\documentclass[preprint,byrevtex, prb,showpacs]{revtex4}
\usepackage{epsfig}
\usepackage{amsmath}
\usepackage{amssymb}
\usepackage{times}
\usepackage{graphicx}

\begin{document}

\title{Study of spin plasmons on the surface of a topological insulator based on spin quantum hydrodynamics}
\author{Ya Zhang$^{1,2}$, Jianwei Cui$^{1}$, Wei Jiang$^{1,3}$, Lin Yi$^{1}$}
\affiliation{$^{1}$  School of Physics, Huazhong University of Science and Technology, Wuhan 430074, China}
\affiliation{$^{2}$  Research group PLASMANT, Department of Chemistry University of Antwerp, B-2610 Wilrijk-Antwerp, Belgium}
\affiliation{$^{3}$  Centre for Mathematical Plasma-Astrophysics, Department of Mathematics, Katholieke Universiteit Leuven, B-3001 Leuven, Belgium}
\email{weijiang@hust.edu.cn}

\begin{abstract}
Starting from the Dirac equation, the relativistic quantum hydrodynamic equations for Dirac electrons on a surface of a three dimensional-topological insulator (TI) are derived and numerically solved to study the spin plasmons. The surface of the TI is modulated by a perpendicular magnetic field and a pulsed exchange field provided by an array of ferromagnetic insulating (FI) stripes. The collective density and velocity oscillations in the two-dimensional Dirac electron gas are simulated and analyzed. It is shown that in spin plasmons, the collective electron density and momentum are modulated by the spin state. Direct experimental observations of such oscillations are possible at laboratory conditions.
\end{abstract}

\maketitle



 Topological insulators (TIs) are found to be a new class of materials which is in insulating states within the bulk, but in conducting edge states on their surfaces \cite{Qi2011}. Three-dimensional (3D) TIs such as Bi$_{2}$Se$_{3}$ shows protected or topological surface state, at which the two-dimensional (2D) electron states can be described as massless 2D Dirac electron gas. Moreover, the 3D TIs are expected to show several unique properties when the time reversal symmetry is broken\cite{Qi2009}. The latter can be achieved directly by a ferromagnetic insulating (FI) layer attached to the 3D TI surface, such that the TI surface states are exchange coupled to the collective magnetization of the FI. The magnetization dynamics of a thin ferromagnetic film exchange coupled with a surface of a 3D-TI has been studied and shown that the ferromagnetic strip could cut the TI surface into two gapless regions\cite{Salehi2011,Tserkovnyak2012}. The collective motion of the Dirac electrons (relativistic spin electrons) are influenced mainly by the magnetization of the FI layer rather than its stray field. This is in contrast to the Schr\"{o}dinger electrons modulated by nanomagnets.

Plasmons are the quantized collective oscillations of electrons in metals and semiconductors. Plasmons can be of appearance under external modulations in varies kind of structures and materials,like bulk metals, semiconductor heterostructures, quantum-wells. Conventional plasmons excited in massive electrons have been extensively studied in plasmonics\cite{Maier2007}. Especially, plasmons of Schr\"{o}dinger electrons which has been a long time research priorities\cite{Pitarke2007}. Like other structured materials, plasmons can also be elicited in the Dirac electron gas, which is often referred as spin plasmons, in which the charge and spin waves are coupled together. Plasmons in Dirac electrons in the TI surfaces has received great attentions since it has been observed \cite{Hsieh2008}, especially in the field of nano-scale electro-mechanical systems. With quantum field-theoretical description, properties and internal structure of spin plasmons in helical liquid are predicted theoretically by Efimkin\cite{Efimkin2012}.  The first experimental observations of spin plasmons are reported by Pietro et al.\cite{Pietro2013} with infrared spectroscopy in thin films on Bi$_{2}$Se$_{3}$ TI surface. Based on Bernevig-Hughes-Zhang model\cite{Bernevig2006}, some theoretical works have studied plasmon excitation dispersion of topological edge states within random-phase approximation (RPA) dielectric theory\cite{Roslyak2013,Juergens2014}, in which the collective motion between Schr\"{o}dinger electrons and Dirac electrons is compared. Within the framework of classical electrodynamics, the dispersion relation of surface spin plasmons at the interface between a TI and a dielectric, has been deduced\cite{Schutky2013}.

However, nearly all above theoretical studies are based on solving the microscopic Dirac-like equations, while the wave function is not but density fluctuations is an observable quantity. It is natural and will be beneficial to adopt a macroscopic theory for the quantum "gas" or "liquid" (also referred as quantum plasma), i.e. a set of hydrodynamic equations, in which the collective density and momentum fluctuations can be directly calculated. The quantum hydrodynamic (QHD) equations  can be deduced from the Schr\"{o}dinger equations, in which the Coulomb interactions can be self-consistently included by coupling the Poisson equation. QHD has been demonstrated as a powerful tools to treat quantum "gas" or "liquid" in high energy density physics\cite{Drake2006} and plasma physics \cite{Shukla2011} community, but is much less familiar for the condensed matter community yet. Compared to solving the nonlinear Schr\"{o}rdinger-Poisson or Wigner-Poisson system, the QHD model is more simple for numerical studies and has a straightforward interpretation in terms of fluid quantities that are employed in classical physics. Spin systems like Dirac electrons on TI surface can also be studied by spin QHD. From Pauli or Dirac equations, hydrodynamic models including non-relativistic spin\cite{Marklund2007} and relativistic spin\cite{Asenjo2011,Tserkovnyak2012}effects have been presented recently. With spin QHD, Felipe\cite{Asenjo2011} has studied electromagnetic wave propagation in spin quantum plasma.

It is possible to use similar treatment to study the spin plasmons, which will be the main work of this paper. We are especially interested in excitations of spin plasmons in Dirac electrons on Bi$_{2}$Se$_{3}$ TI surface with FI strips and a perpendicular magnetic field. The Bi$_{2}$Se$_{3}$ material has been verified\cite{Xia2009} to have a bulk gap of $0.3$ eV and a single Dirac cone of surface states. The collective motion under these external modulation and a spin force of the Dirac electrons can behave as a tool of surface modification of the TI, or as a useful probe to characterize the average electromagnetic properties of the 2D Dirac electron system. The works have two goals, first we will deduce the spin QHD suitable for describe the spin plasmons starting from Dirac equation with similar procedures in Ref.\cite{Marklund2007}, after that we will study the collective properties of the spin plasmons in such a system numerically. Gauss units will be adopted throughout the paper except in specific definitions.

We consider a 2D semi-infinite Dirac electron gas on an identified surface of a 3D TI like Bi$_{2}$Se$_{3}$ material ($v_{F}=5\times 10^{7}$ cm/s) with initial density of $n_{0}=3\times 10^{14}$ cm$^{-2}$, in which $n_{0}=3 k_{F}^{2}/2\pi$ and $k_{F}=m_{e}v_{F}/\hbar$. Here $v_{F}$ is the Fermi velocity of the electrons, $k_{F}$ is the Fermi wave number, $m_{e}$ is the electron mass, $\hbar$ is the Planck constant divided by $2\pi$.
Take a cartesian coordinate system $\{z ,x\}$ in the surface and the 2D surface electrons are in the region $x \geq 0$.
The in-plane FI stripes are deposited periodically on top of the surface along the $x$ axis, as sketched in Fig. 1.
Their initial magnetizations are along the $z$ axis in the $\{z ,x\}$ plane. All FI stripes take the same width $d/2$ and magnetization strength $m_{0}$, the
smallest distance between them is $d/2$ and the periodic length is $d=5$ nm.

In addition, an external perpendicular-plane TE wave with the electromagnetic fields $\mathbf{E}=(E_{x},0,E_{z})$ and $\mathbf{B}=(0,B_{y},0)$, is applied to the surface which is propagating along the $x$ axis and directed along the $y$ axis with $k^{2}=k_{x}^{2}$, in the analytical formulation of
\begin{equation}
\begin{array}{lcl}
    E_{x}(x,t)=B_{0}\cos(k_{x}x),  \\
    B_{y}(x,t)=-B_{0}k\cos(k_{x}x)\cos(kct)/k_{x}\\
    E_{z}(x,t)=B_{0}\sin(k_{x}x)\sin(kct)
 \end{array}
\text{.}  \label{B}
\end{equation}%
The boundary condition at $x=0$ can be expressed as a perfectly conducting wall.

In the presence of the perpendicular-plane TE wave and in-plane FI stripes modulations, the motion of surface electrons that take collective polarization can be regarded as relativistic surface plasmons. The plasmons which are regarded as charged fluids with 2D average density field $n_{e}(z,x,t)$ and velocity field $\mathbf{u}_{e}(z,x,t)$ can be described by the relativistic QHD equations. Now comes the crucial point: by introducing the decomposition of the spinors according to $\psi=\sqrt{\gamma n}\exp(iS/\hbar)\varphi$ with the relation $m_{e}\mathbf{u}_{e}=\mathbf{\nabla }S+e\mathbf{A}/c$,
it is possible to derive relativistic quantum continuity and momentum equations from the EOM of relativistic fluid and the Dirac equation
\begin{equation}
i\hbar\frac{\partial \psi}{\partial t}=v_{F}\mathbf{\sigma}\cdot(\mathbf{p}-\frac{e}{c}\mathbf{A})\psi-e\phi\psi+\mathbf{\sigma}\cdot \mathbf{M}\psi.  \label{H}
\end{equation}%
Here $\gamma=1/\sqrt{1-u_{e}^{2}/v_{F}^{2}}$ is the relativistic factor, $e$ is the elementary charge and $c$ is the light speed,
$\mathbf{p}=(p_{z} , p_{x})$ is the electron momentum, $\mathbf{\sigma}=(\sigma_{x},\sigma_{y},\sigma_{z})$ are the Pauli spin matrices, and $\varphi$ is the 2-spinor with $\varphi\varphi^{+}=1$. Beside, $\phi=\phi _{ext}+\phi _{ind}$ is the total potential consisted of the external potential $\phi _{ext}$ related to the $E_{x}$ and $E_{z}$ and the induced electrostatic potential $\phi_{ind}$ generated by the collective motion of the electrons in an electrostatic case considered here.
In particular, $\mathbf{M}
=m_{z}(x)\mathbf{e}_{z}$ is the effective exchange field induced from the FI stripes, and $m_{z}(x)$ takes a constant value $m_{0}$ in the stripe regions with magnetization
aligned to the $z$ axis and zero otherwise. Moreover, the last term in the right side of Eq. (\ref{H}) is induced by the spin effect where the magnitude of $\mathbf{M}$, $m_{0}$ (can also be defined as the value of the exchange field), represents the strength of the spin effect.
Thus, the collective motions of the surface Dirac electrons can be described by the continuity equation
\begin{equation}
\frac{\partial \gamma n_{e}}{\partial t}+\mathbf{\nabla }\cdot \left(
\gamma n_{e}\mathbf{u}_{e}\right) =0,  \label{num}
\end{equation}%
and the momentum-balance equation
\begin{align}
& \frac{\partial \mathbf{u}_{e}}{\partial t}+\left( \mathbf{u}_{e}\cdot
\mathbf{\nabla }\right) \mathbf{u}_{e}=\frac{e}{\gamma^{3} m_{e}}(\mathbf{\nabla }\phi _{ext}+\mathbf{\nabla }\phi _{ind}+\frac{1}{c}\frac{\partial \mathbf{A}}{\partial t}-\frac{\mathbf{u}_{e}\times\mathbf{B}}{c})\mp\frac{1}{\gamma m_{e}}\mathbf{\nabla }m_{z}.  \label{mom}
\end{align}%
Here, "$-$" represents spin-up and "$+$" represents spin-down in the last term on the right side of Eq. (\ref{mom}) that is the spin force term, $\mathbf{\nabla }=\frac{\partial }{\partial z}\mathbf{e}%
_{z}+\frac{\partial }{\partial x}\mathbf{e}_{x}$,
and the magnetic vector potential $\mathbf{A}=(0,0,A_{z})$  can be given from Eq. (\ref{B}) as
\begin{equation}
     A_{z}(x,t)=B_{0}k\sin(k_{x}x)\cos(kct)/k_{x}^{2}
\text{.}\label{maxwell}
\end{equation}
Note in this momentum equation, no dissipating term is included.
The induced electrostatic potential $\phi _{ind}$ satisfies the Poisson equation
\begin{equation}
\nabla ^{2}\phi _{ind}=4\pi e(n_{e}-n_{0}).  \label{poisson}
\end{equation}

Above equations are nonlinearly coupled and can only solved numerically. The Poisson equation (\ref{poisson}) is solved by the successive over relaxation (SOR) method. Flux-corrected transport (FCT) method\cite{Boris1993} is adopted to numerically solve the Eqs. (\ref{num}) and (\ref{mom}) by time integration from the initial time $t=0$ when the values of all quantities are known.

For convenience we introduce the length of a basic unit $a=5\times10^{-8}$ cm. Note that we examine the collective polarization of 2D Dirac electrons on the surface of Bi$_{2}$Se$_{3}$ TI material
in the external TE wave and the FI strip modulations and by considering spin effects in the following results. In our simulations, we take several period length of both $B_{y}$ and $m_{z}$ in $x$ direction ($x/a=0-32$) as simulation region,
the initial electron density $n_{0}=3\times 10^{14}$ cm$^{-2}$ and velocity $\mathbf{u}_{e}=0$ are treated as given parameters, while the values of $m_{0}$, $B_{0}$ and the wave number $k$
are varied in order to examine these external modulations on the collective motion of Dirac electrons.

The spin effects on the collective density oscillations at 1.5 fs are illustrated in Fig. 2 without spin effects $m_{0}=0$ (solid line) and with spin effects $m_{0}=1$ meV ($\sim17$ T) (spin-up: dashed line, spin-down: dotted line). Laboratory magnetic fields are at most several tens of T.
Here $B_{0}=0.1$ T for $ka=\pi/8$ (a), $ka=\pi/4$ (b), $ka=\pi/2$ (c) and $ka=\pi$ (d).
It is found from Fig. 2 that the collective oscillations with spin effects are larger than that without spin effects, showing as periodical spikes in the charge density. It can be clearly seen that, the spin state is locked with the density waves, which is mostly obvious when $ka=\pi$. Different spin state will lead to different sign of density oscillations. We can also note that the contribution from the spin term is significant in particular for large values of the wave number $k$. In addition, it is interesting to see that the spin-up effects are more significant when $ka=\pi/2$ than spin-down effects.
Note for most classical plasmas, the strength of magnetization is very small while the temperature is high, the spins mainly randomly orient, and the spin
effects are negligible. On the other hand, when there exist strongly magnetic modulations (larger than several T) in a quantum plasma, the spin vectors are essentially modulated and the spin effects can be significant.
Thus, in our calculation, when the exchange fields induced by the FI stripes take a strength of several meV or several tens of T, spin effects are significant, which can meet the most of the experimental conditions. That is, the spin effects are evident on the collective motion of Dirac electrons, leading to the increase of the density oscillation.

As demonstrated above, the collective polarization features with ($m_{0}=1$ meV) and without ($m_{0}=0$) spin effects are quite distinct. In addition, the different wave number $k$ of the perpendicular-plane magnetic field $B_{y}$ also influences the collective motion of the electrons.  Thus, some interesting questions arise: How are the density oscillations changed with and without the external magnetic field $B_{y}$ driving them, what happens if at the time $t_{B}$ the $B_{y}$ is turned off, and how are the plasma oscillations sustained when $B_{y}$ is turned off at the time $t_{B}$? Here $t_{B}$ is the time when the $B_{y}$ is turned off. In order to answer these questions, Fig. 3 presents the density oscillations at $3$ fs: (a-c) with $B_{y}$ ($B_{0}=0.1$ T and $ka=\pi/4$), $t_{B}=3$ fs (solid line), $t_{B}=1.5$ fs (dashed line), for non-spin (a), spin-up (b) and spin-down (c), and (d) without $B_{y}$ for spin-up (square symbol) and spin-down (circle symbol). Here $m_{0}=1$ meV.
In Fig. 3(a) without spin effect, the density oscillations decrease obviously when the $B_{y}$ is turned off but the oscillations persist by the waked potential [see Eq. (\ref{poisson})]. However, in Fig. 3 (b-c) with spin effects the density oscillations show a much smaller change when the $B_{y}$ is turned off (dashed line) compared to the case with $B_{y}$ (solid line). In particular, the density oscillations can be sustained with spin effects even there is no $B_{y}$ ($B_{0}$=0), as shown in Fig. 3(d).
That is, a perpendicular-magnetic field of a few T ($B_{0}=0.1$ T), as well as an in-plane exchange field of several meV ($m_{0}=1$ meV) which appears in the spin force term in Eq. (\ref{mom}), both might generate and dominate the density oscillations.

In order to see the collective oscillations of the 2D Dirac electrons more clearly, Fig. 4 shows the spin-up effects on the collective density oscillations (a), longitudinal ($x$ direction) velocity $u_{ex}$ (b) and transverse ($z$ direction) velocity $u_{ez}$ (c) at 3 fs in 2D spectrums. Here $ka=\pi/4$, $B_{0}=0.1$ T and $m_{0}=1$ meV. The velocity is normalized by $v_{F}$. It is easy to see that the density oscillates with respect to the $x$ axis and the maximum positive and negative densities are intersected on both sides of the $x$ axis in the 2D spectrum, as shown in Fig. 4(a). Whereas, the peak values of the both longitudinal and transverse velocities are on the $x$ axis due to the maximum force terms on the $x$ axis, as shown in Fig. 4(b-c). What is interesting, $u_{ex}$ is much larger than $u_{ex}$ due to the spin-momentum locking effect. This is the spin is in $z$ direction and enhance the momentum of $x$ direction.

In conclusion, we have studied the collective oscillations of Dirac electrons on the surface of a 3D-TI subject to a perpendicular-plane TE wave and an in-plane periodic exchange field which is provided by a series of equally spaced FI stripes attached to the TI surface. The collective density and velocity of this electrons are obtained by the relativistic quantum hydrodynamic equations of continuity and momentum which are coupled with the Poisson equation solved by numerically. In our simulation, as the surface electron gas of a 3D-TI is considered, the spin effects are included naturally and they are significant under the spin force assisted by the in-plane FI stripes modulation with a laboratory condition, especially the spin effects can sustain the density oscillation when the TE wave is turned off. In addition, the TE wave modulation is also obvious for different wave numbers. In a 2D spectrum, the collective oscillations of the 2D Dirac electrons are clearly seen, in which the longitudinal velocity is much larger than the transverse velocity due to the spin-momentum locking effect. The simulation results will likely find its experimental application in relativistic quantum plasmas including Dirac electron plasmas and solid density plasmas, especially in the research of the surface states of 3D-TIs.

Here we should emphasis that it was generally believed by the plasma physics community that the spin quantum hydrodynamics related effects can be never be observed in experiments\cite{Marklund2007}, as it may require the external magnetic field as high as several thousand T in 3D bulk materials, like in a neutral star. However, our work shows that in a reduced 2D Dirac electron system on the surface of a TI, the spin effects on the collective motion of these electrons can be observed with only several tens of T, showing another significant example of the low-dimensional phenomenon.

The authors thanks the referees for several helpful suggestions. W. Jiang and Y. Zhang gratefully acknowledge the Belgian Federal Science Policy Office and the China Scholarship Council for financial support. This work was partially supported by the NSFC (11105057, 11275007).

\bibliography{TI}

\begin{thebibliography}{19}
\expandafter\ifx\csname natexlab\endcsname\relax\def\natexlab#1{#1}\fi
\expandafter\ifx\csname bibnamefont\endcsname\relax
  \def\bibnamefont#1{#1}\fi
\expandafter\ifx\csname bibfnamefont\endcsname\relax
  \def\bibfnamefont#1{#1}\fi
\expandafter\ifx\csname citenamefont\endcsname\relax
  \def\citenamefont#1{#1}\fi
\expandafter\ifx\csname url\endcsname\relax
  \def\url#1{\texttt{#1}}\fi
\expandafter\ifx\csname urlprefix\endcsname\relax\def\urlprefix{URL }\fi
\providecommand{\bibinfo}[2]{#2}
\providecommand{\eprint}[2][]{\url{#2}}

\bibitem[{\citenamefont{Qi and Zhang}(2011)}]{Qi2011}
\bibinfo{author}{\bibfnamefont{X.-L.} \bibnamefont{Qi}} \bibnamefont{and}
  \bibinfo{author}{\bibfnamefont{S.-C.} \bibnamefont{Zhang}},
  \bibinfo{journal}{Reviews of Modern Physics} \textbf{\bibinfo{volume}{83}},
  \bibinfo{pages}{1057} (\bibinfo{year}{2011}).

\bibitem[{\citenamefont{Qi et~al.}(2009)\citenamefont{Qi, Li, Zang, and
  Zhang}}]{Qi2009}
\bibinfo{author}{\bibfnamefont{X.-L.} \bibnamefont{Qi}},
  \bibinfo{author}{\bibfnamefont{R.}~\bibnamefont{Li}},
  \bibinfo{author}{\bibfnamefont{J.}~\bibnamefont{Zang}}, \bibnamefont{and}
  \bibinfo{author}{\bibfnamefont{S.-C.} \bibnamefont{Zhang}},
  \bibinfo{journal}{Science} \textbf{\bibinfo{volume}{323}},
  \bibinfo{pages}{1184} (\bibinfo{year}{2009}).

\bibitem[{\citenamefont{Salehi et~al.}(2011)\citenamefont{Salehi, Alidoust,
  Rahnavard, and Rashedi}}]{Salehi2011}
\bibinfo{author}{\bibfnamefont{M.}~\bibnamefont{Salehi}},
  \bibinfo{author}{\bibfnamefont{M.}~\bibnamefont{Alidoust}},
  \bibinfo{author}{\bibfnamefont{Y.}~\bibnamefont{Rahnavard}},
  \bibnamefont{and} \bibinfo{author}{\bibfnamefont{G.}~\bibnamefont{Rashedi}},
  \bibinfo{journal}{Physica E: Low-dimensional Systems and Nanostructures}
  \textbf{\bibinfo{volume}{43}}, \bibinfo{pages}{966} (\bibinfo{year}{2011}).

\bibitem[{\citenamefont{Tserkovnyak and Loss}(2012)}]{Tserkovnyak2012}
\bibinfo{author}{\bibfnamefont{Y.}~\bibnamefont{Tserkovnyak}} \bibnamefont{and}
  \bibinfo{author}{\bibfnamefont{D.}~\bibnamefont{Loss}},
  \bibinfo{journal}{Phys. Rev. Lett.} \textbf{\bibinfo{volume}{108}},
  \bibinfo{pages}{187201} (\bibinfo{year}{2012}),
  \urlprefix\url{http://link.aps.org/doi/10.1103/PhysRevLett.108.187201}.

\bibitem[{\citenamefont{Maier}(2007)}]{Maier2007}
\bibinfo{author}{\bibfnamefont{S.~A.} \bibnamefont{Maier}},
  \emph{\bibinfo{title}{Plasmonics: Fundamentals and Applications: Fundamentals
  and Applications}} (\bibinfo{publisher}{Springer}, \bibinfo{year}{2007}).

\bibitem[{\citenamefont{Pitarke et~al.}(2007)\citenamefont{Pitarke, Silkin,
  Chulkov, and Echenique}}]{Pitarke2007}
\bibinfo{author}{\bibfnamefont{J.~M.} \bibnamefont{Pitarke}},
  \bibinfo{author}{\bibfnamefont{V.~M.} \bibnamefont{Silkin}},
  \bibinfo{author}{\bibfnamefont{E.~V.} \bibnamefont{Chulkov}},
  \bibnamefont{and} \bibinfo{author}{\bibfnamefont{P.~M.}
  \bibnamefont{Echenique}}, \bibinfo{journal}{Reports on Progress in Physics}
  \textbf{\bibinfo{volume}{70}}, \bibinfo{pages}{1} (\bibinfo{year}{2007}),
  \urlprefix\url{http://stacks.iop.org/0034-4885/70/i=1/a=R01}.

\bibitem[{\citenamefont{Hsieh et~al.}(2008)\citenamefont{Hsieh, Qian, Wray,
  Xia, Hor, Cava, and Hasan}}]{Hsieh2008}
\bibinfo{author}{\bibfnamefont{D.}~\bibnamefont{Hsieh}},
  \bibinfo{author}{\bibfnamefont{D.}~\bibnamefont{Qian}},
  \bibinfo{author}{\bibfnamefont{L.}~\bibnamefont{Wray}},
  \bibinfo{author}{\bibfnamefont{Y.}~\bibnamefont{Xia}},
  \bibinfo{author}{\bibfnamefont{Y.~S.} \bibnamefont{Hor}},
  \bibinfo{author}{\bibfnamefont{R.}~\bibnamefont{Cava}}, \bibnamefont{and}
  \bibinfo{author}{\bibfnamefont{M.~Z.} \bibnamefont{Hasan}},
  \bibinfo{journal}{Nature} \textbf{\bibinfo{volume}{452}},
  \bibinfo{pages}{970} (\bibinfo{year}{2008}).

\bibitem[{\citenamefont{Efimkin et~al.}(2012)\citenamefont{Efimkin, Lozovik,
  and Sokolik}}]{Efimkin2012}
\bibinfo{author}{\bibfnamefont{D.~K.} \bibnamefont{Efimkin}},
  \bibinfo{author}{\bibfnamefont{Y.~E.} \bibnamefont{Lozovik}},
  \bibnamefont{and} \bibinfo{author}{\bibfnamefont{A.~A.}
  \bibnamefont{Sokolik}}, \bibinfo{journal}{Nanoscale research letters}
  \textbf{\bibinfo{volume}{7}}, \bibinfo{pages}{1} (\bibinfo{year}{2012}).

\bibitem[{\citenamefont{Di~Pietro et~al.}(2013)\citenamefont{Di~Pietro,
  Ortolani, Limaj, Di~Gaspare, Giliberti, Giorgianni, Brahlek, Bansal, Koirala,
  Oh et~al.}}]{Pietro2013}
\bibinfo{author}{\bibfnamefont{P.}~\bibnamefont{Di~Pietro}},
  \bibinfo{author}{\bibfnamefont{M.}~\bibnamefont{Ortolani}},
  \bibinfo{author}{\bibfnamefont{O.}~\bibnamefont{Limaj}},
  \bibinfo{author}{\bibfnamefont{A.}~\bibnamefont{Di~Gaspare}},
  \bibinfo{author}{\bibfnamefont{V.}~\bibnamefont{Giliberti}},
  \bibinfo{author}{\bibfnamefont{F.}~\bibnamefont{Giorgianni}},
  \bibinfo{author}{\bibfnamefont{M.}~\bibnamefont{Brahlek}},
  \bibinfo{author}{\bibfnamefont{N.}~\bibnamefont{Bansal}},
  \bibinfo{author}{\bibfnamefont{N.}~\bibnamefont{Koirala}},
  \bibinfo{author}{\bibfnamefont{S.}~\bibnamefont{Oh}}, \bibnamefont{et~al.},
  \bibinfo{journal}{Nature nanotechnology} \textbf{\bibinfo{volume}{8}},
  \bibinfo{pages}{556} (\bibinfo{year}{2013}).

\bibitem[{\citenamefont{Bernevig et~al.}(2006)\citenamefont{Bernevig, Hughes,
  and Zhang}}]{Bernevig2006}
\bibinfo{author}{\bibfnamefont{B.~A.} \bibnamefont{Bernevig}},
  \bibinfo{author}{\bibfnamefont{T.~L.} \bibnamefont{Hughes}},
  \bibnamefont{and} \bibinfo{author}{\bibfnamefont{S.-C.} \bibnamefont{Zhang}},
  \bibinfo{journal}{Science} \textbf{\bibinfo{volume}{314}},
  \bibinfo{pages}{1757} (\bibinfo{year}{2006}).

\bibitem[{\citenamefont{Roslyak et~al.}(2013)\citenamefont{Roslyak, Gumbs, and
  Huang}}]{Roslyak2013}
\bibinfo{author}{\bibfnamefont{O.}~\bibnamefont{Roslyak}},
  \bibinfo{author}{\bibfnamefont{G.}~\bibnamefont{Gumbs}}, \bibnamefont{and}
  \bibinfo{author}{\bibfnamefont{D.}~\bibnamefont{Huang}},
  \bibinfo{journal}{Phys. Rev. B} \textbf{\bibinfo{volume}{87}},
  \bibinfo{pages}{045121} (\bibinfo{year}{2013}),
  \urlprefix\url{http://link.aps.org/doi/10.1103/PhysRevB.87.045121}.

\bibitem[{\citenamefont{Juergens et~al.}(2014)\citenamefont{Juergens, Michetti,
  and Trauzettel}}]{Juergens2014}
\bibinfo{author}{\bibfnamefont{S.}~\bibnamefont{Juergens}},
  \bibinfo{author}{\bibfnamefont{P.}~\bibnamefont{Michetti}}, \bibnamefont{and}
  \bibinfo{author}{\bibfnamefont{B.}~\bibnamefont{Trauzettel}},
  \bibinfo{journal}{Phys. Rev. Lett.} \textbf{\bibinfo{volume}{112}},
  \bibinfo{pages}{076804} (\bibinfo{year}{2014}),
  \urlprefix\url{http://link.aps.org/doi/10.1103/PhysRevLett.112.076804}.

\bibitem[{\citenamefont{Sch\"utky et~al.}(2013)\citenamefont{Sch\"utky, Ertler,
  Tr\"ugler, and Hohenester}}]{Schutky2013}
\bibinfo{author}{\bibfnamefont{R.}~\bibnamefont{Sch\"utky}},
  \bibinfo{author}{\bibfnamefont{C.}~\bibnamefont{Ertler}},
  \bibinfo{author}{\bibfnamefont{A.}~\bibnamefont{Tr\"ugler}},
  \bibnamefont{and}
  \bibinfo{author}{\bibfnamefont{U.}~\bibnamefont{Hohenester}},
  \bibinfo{journal}{Phys. Rev. B} \textbf{\bibinfo{volume}{88}},
  \bibinfo{pages}{195311} (\bibinfo{year}{2013}),
  \urlprefix\url{http://link.aps.org/doi/10.1103/PhysRevB.88.195311}.

\bibitem[{\citenamefont{Drake}(2006)}]{Drake2006}
\bibinfo{author}{\bibfnamefont{R.~P.} \bibnamefont{Drake}},
  \emph{\bibinfo{title}{High-energy-density physics: fundamentals, inertial
  fusion, and experimental astrophysics}} (\bibinfo{publisher}{Springer},
  \bibinfo{year}{2006}).

\bibitem[{\citenamefont{Shukla and Eliasson}(2011)}]{Shukla2011}
\bibinfo{author}{\bibfnamefont{P.}~\bibnamefont{Shukla}} \bibnamefont{and}
  \bibinfo{author}{\bibfnamefont{B.}~\bibnamefont{Eliasson}},
  \bibinfo{journal}{Reviews of Modern Physics} \textbf{\bibinfo{volume}{83}},
  \bibinfo{pages}{885} (\bibinfo{year}{2011}).

\bibitem[{\citenamefont{Marklund and Brodin}(2007)}]{Marklund2007}
\bibinfo{author}{\bibfnamefont{M.}~\bibnamefont{Marklund}} \bibnamefont{and}
  \bibinfo{author}{\bibfnamefont{G.}~\bibnamefont{Brodin}},
  \bibinfo{journal}{Physical review letters} \textbf{\bibinfo{volume}{98}},
  \bibinfo{pages}{025001} (\bibinfo{year}{2007}).

\bibitem[{\citenamefont{Asenjo et~al.}(2011)\citenamefont{Asenjo, Mu{\~n}oz,
  Valdivia, and Mahajan}}]{Asenjo2011}
\bibinfo{author}{\bibfnamefont{F.~A.} \bibnamefont{Asenjo}},
  \bibinfo{author}{\bibfnamefont{V.}~\bibnamefont{Mu{\~n}oz}},
  \bibinfo{author}{\bibfnamefont{J.~A.} \bibnamefont{Valdivia}},
  \bibnamefont{and} \bibinfo{author}{\bibfnamefont{S.~M.}
  \bibnamefont{Mahajan}}, \bibinfo{journal}{Physics of Plasmas (1994-present)}
  \textbf{\bibinfo{volume}{18}}, \bibinfo{pages}{012107}
  (\bibinfo{year}{2011}).

\bibitem[{\citenamefont{Xia et~al.}(2009)\citenamefont{Xia, Qian, Hsieh, Wray,
  Pal, Lin, Bansil, Grauer, Hor, Cava et~al.}}]{Xia2009}
\bibinfo{author}{\bibfnamefont{Y.}~\bibnamefont{Xia}},
  \bibinfo{author}{\bibfnamefont{D.}~\bibnamefont{Qian}},
  \bibinfo{author}{\bibfnamefont{D.}~\bibnamefont{Hsieh}},
  \bibinfo{author}{\bibfnamefont{L.}~\bibnamefont{Wray}},
  \bibinfo{author}{\bibfnamefont{A.}~\bibnamefont{Pal}},
  \bibinfo{author}{\bibfnamefont{H.}~\bibnamefont{Lin}},
  \bibinfo{author}{\bibfnamefont{A.}~\bibnamefont{Bansil}},
  \bibinfo{author}{\bibfnamefont{D.}~\bibnamefont{Grauer}},
  \bibinfo{author}{\bibfnamefont{Y.}~\bibnamefont{Hor}},
  \bibinfo{author}{\bibfnamefont{R.}~\bibnamefont{Cava}}, \bibnamefont{et~al.},
  \bibinfo{journal}{Nature Physics} \textbf{\bibinfo{volume}{5}},
  \bibinfo{pages}{398} (\bibinfo{year}{2009}).

\bibitem[{\citenamefont{Boris et~al.}(1993)\citenamefont{Boris, Landsberg,
  Oran, and Gardner}}]{Boris1993}
\bibinfo{author}{\bibfnamefont{J.~P.} \bibnamefont{Boris}},
  \bibinfo{author}{\bibfnamefont{A.~M.} \bibnamefont{Landsberg}},
  \bibinfo{author}{\bibfnamefont{E.~S.} \bibnamefont{Oran}}, \bibnamefont{and}
  \bibinfo{author}{\bibfnamefont{J.~H.} \bibnamefont{Gardner}},
  \bibinfo{type}{Tech. Rep.}, \bibinfo{institution}{DTIC Document}
  (\bibinfo{year}{1993}).

\end{thebibliography}
\newpage \hfill

\begin{center}
\textbf{Figure Captions}
\end{center}

\hfill

\begin{figure}[htbp]
\begin{center}
\includegraphics[height=5.0cm,width=7.0cm]{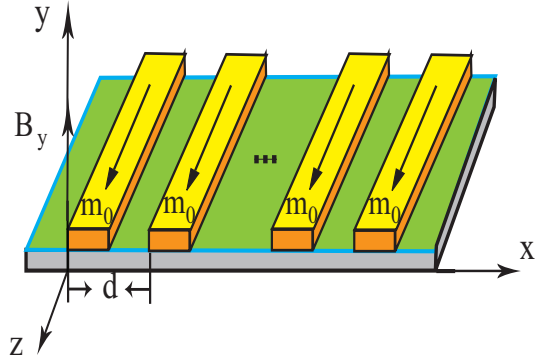}
\end{center}
\caption{Schematic illustration of the device: the 2D Dirac electrons in the
surface of a TI is modulated by a pulsed exchange field. The latter is
generated by an array of FI stripes. The magnetization directions of
FI stripes are parallel to the $z$ axis. Both
the FI regions and the spacing regions have the same length $d/2$($d=5$ nm). A perpendicular
TE wave $B_{y}$ propagating in the surface of the TI along the $x$ direction.}
\label{fig:Fig1}
\end{figure}

\hfill

\begin{figure}[htbp]
\begin{center}
\includegraphics[height=10.0cm,width=12.0cm]{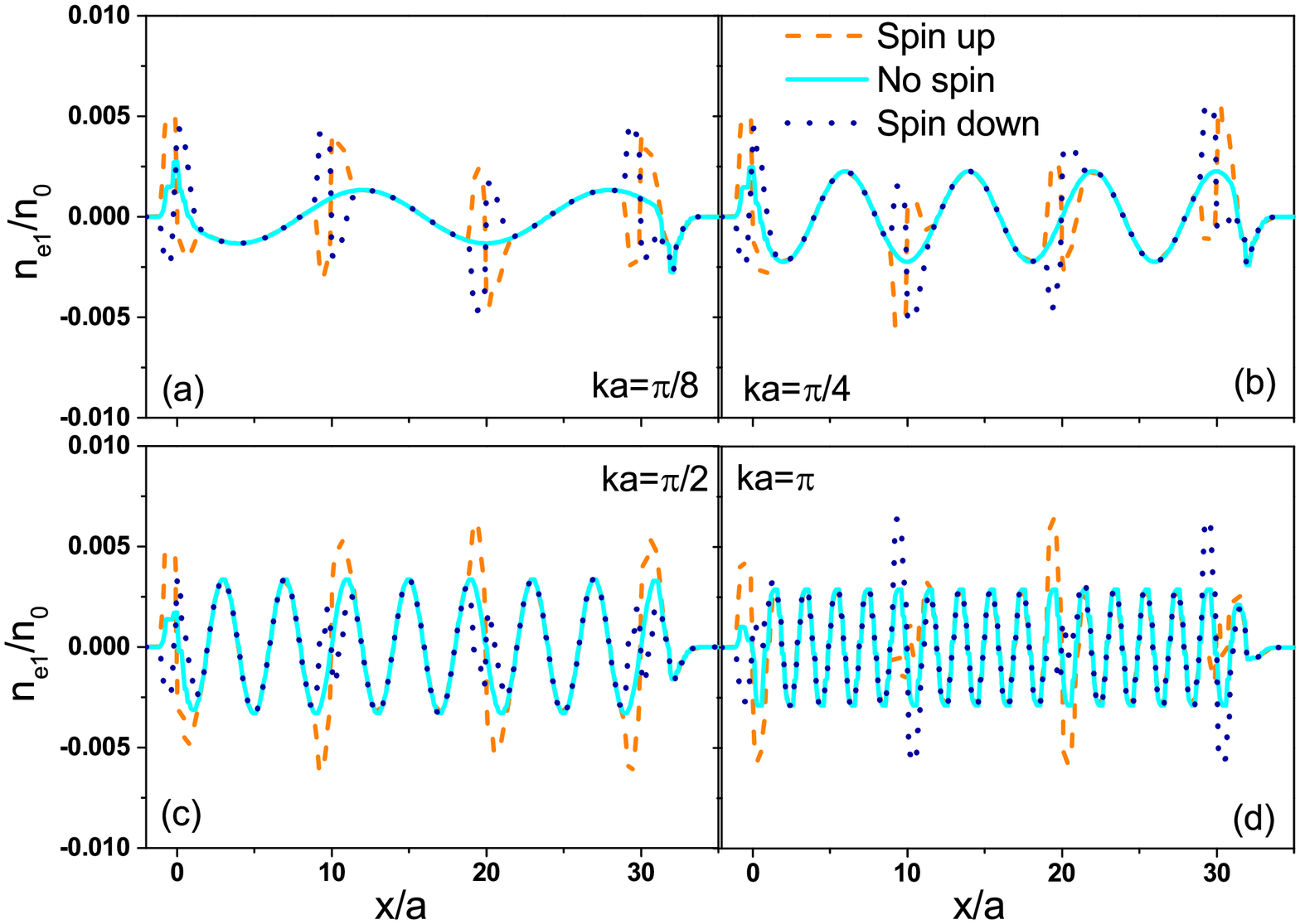}
\end{center}
\caption{(Color online) The density oscillation at 1.5 fs ($n_{e1}=n_{e}-n_{0}$) versus $x$
for (a-d) $B_{0}=0.1$ T with (a) $ka=\pi/8$, (b) $ka=\pi/4$, (c) $ka=\pi/2$ and (d) $ka=\pi$, without spin effects $m_{0}=0$ (solid line) and with spin effects $m_{0}=1$ meV (spin-up: dashed line) and (spin-down: dotted line). }
\label{fig:Fig2}
\end{figure}

\hfill

\begin{figure}[tbph]
\begin{center}
\includegraphics[height=10.0cm,width=12.0cm]{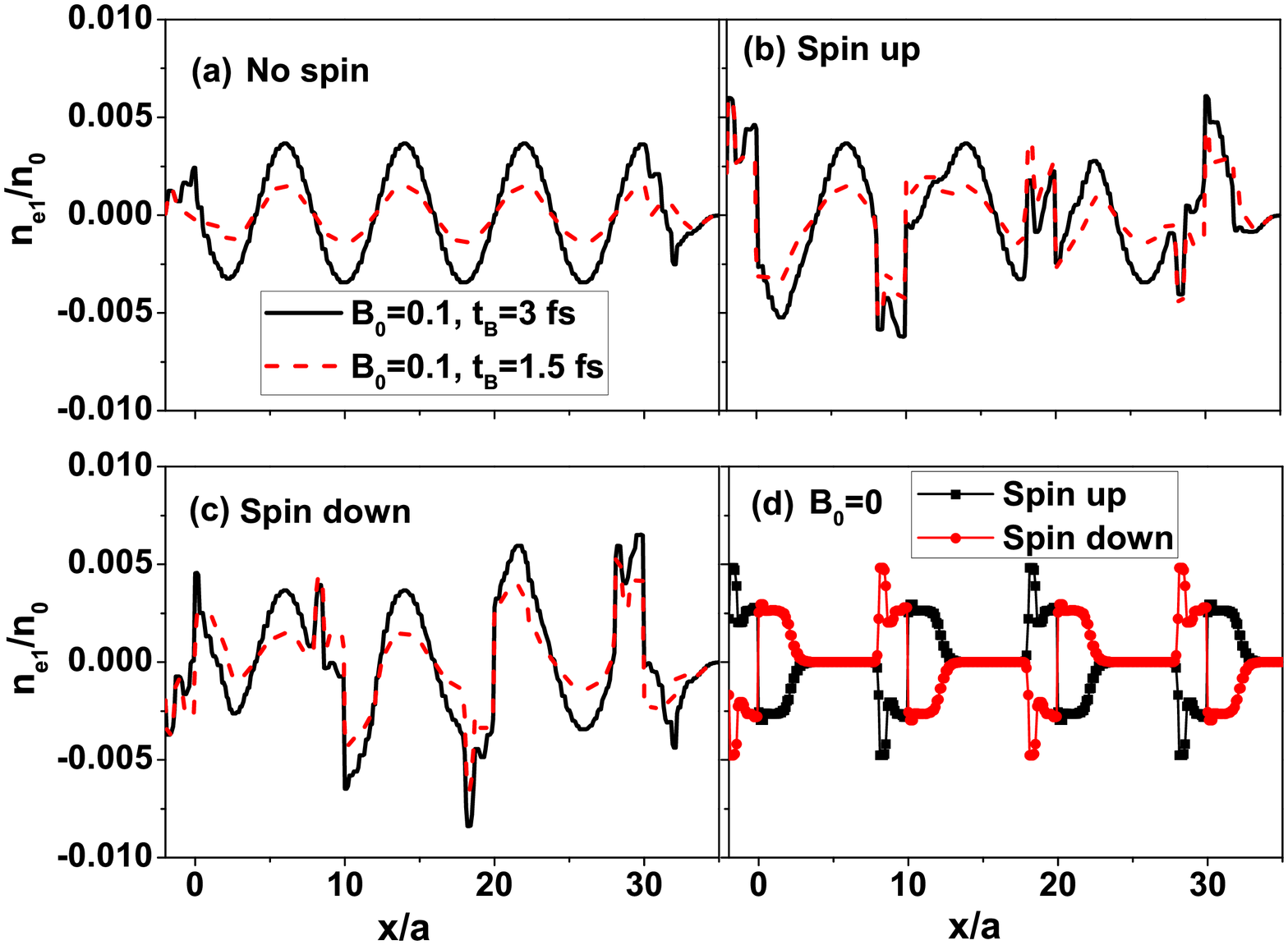}
\end{center}
\caption{(Color online) The density oscillations at $3$ fs: (a-c) with $B_{y}$ ($B_{0}=0.1$ T and $ka=\pi/4$), $t_{B}=3$ fs (solid line) and $t_{B}=1.5$ fs (dashed line), for non-spin (a), spin-up (b) and spin-down (c), and (d) without $B_{y}$ for spin-up (square symbol) and spin-down (circle symbol). $m_{0}=1$ meV. Here $t_{B}$ is the time when the $B_{y}$ is turned off. }
\label{fig:Fig3}
\end{figure}

\hfill
\begin{figure}[tbph]
\begin{center}
\includegraphics[height=8.0cm,width=10.0cm]{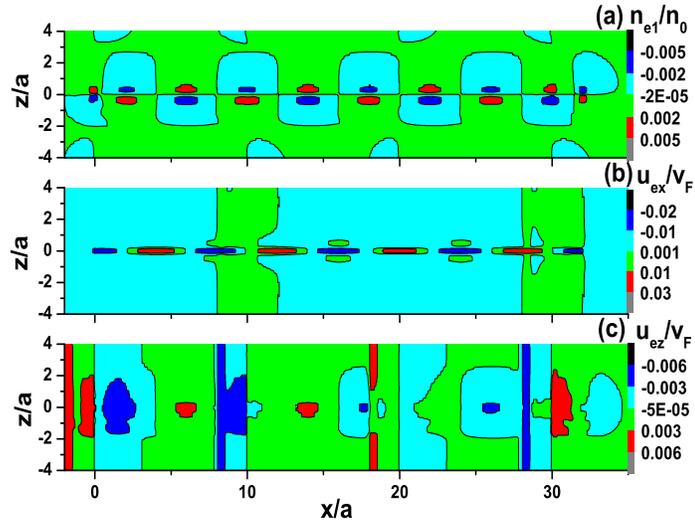}
\end{center}
\caption{(Color online) The spin-up effects on the density oscillation ($n_{e1}=n_{e}-n_{0}$) (a), longitudinal velocity $u_{ex}$ (b) and transverse velocity $u_{ez}$ (c) at 3 fs in a 2D spectrum.
Here $ka=\pi/4$, $B_{0}=0.1$ T and $m_{0}=1$ meV.}
\label{fig:Fig4}
\end{figure}

\end{document}